\documentclass[aps,prb,twocolumn,superscriptaddress,showpacs,showkeys,floatfix]{revtex4-1}

\usepackage{graphicx}
\usepackage{amsmath}
\usepackage{amssymb}

\begin{document}


\title{The thermal denaturation of DNA studied with neutron scattering}

\author{Andrew Wildes}
\affiliation{Institut Laue Langevin,BP 156, 6, rue Jules Horowitz
38042 Grenoble Cedex 9, France}

\author{Nikos Theodorakopoulos}
\affiliation{Theoretical and Physical Chemistry Institute, NHRF, 116 35 Athens, Greece}
\affiliation{Fachbereich Physik, Universit\"at Konstanz, 78457
  Konstanz, Germany} 

\author{Jessica Valle-Orero}
\affiliation{Institut Laue Langevin,BP 156, 6, rue Jules Horowitz
38042 Grenoble Cedex 9, France}
\affiliation{Ecole 
Normale Sup\'erieure de Lyon, Laboratoire de Physique CNRS,
46 all\'ee d'Italie, 69364 Lyon Cedex 7, France}

\author{Santiago Cuesta-L\'opez}
\affiliation{Ecole 
Normale Sup\'erieure de Lyon, Laboratoire de Physique CNRS,
46 all\'ee d'Italie, 69364 Lyon Cedex 7, France}

\author{Jean-Luc Garden}
\affiliation{Institut N\'eel, CNRS - Universit\'e Joseph Fourier,
BP 166, 38042 Grenoble cedex 9, France.}

\author{Michel Peyrard}
\affiliation{Ecole 
Normale Sup\'erieure de Lyon, Laboratoire de Physique CNRS,
46 all\'ee d'Italie, 69364 Lyon Cedex 7, France}


\date{\today}

\begin{abstract}
The melting transition of deoxyribonucleic acid (DNA), whereby the strands of the double helix structure completely separate at a certain temperature, has been characterized using neutron scattering.  A Bragg peak from B-form fibre DNA has been measured as a function of temperature, and its widths and integrated intensities have been interpreted using the Peyrard-Bishop-Dauxois (PBD) model with only one free parameter.  The experiment is unique, as it gives spatial correlation along the molecule through the melting transition where other techniques cannot.
\end{abstract}


\maketitle
Deoxyribonucleic acid (DNA) is a highly dynamic molecule in which the base pairs, which carry the genetic information, fluctuate widely.  This can lead to a temporary breaking of a ``closed" pair and a local separation of the two strands.  Local openings may be activated by heating.  At a certain temperature, local openings of the double helix extend over the full molecule, resulting in a complete separation of the two strands.  This is called the ``melting'' of DNA, and may be considered a very rare example of a one-dimensional (1D) structural phase transition \cite{Zimm,Dauxois}.   DNA melting attracted attention soon after the discovery of the double helix structure \cite{Thomas,Rice} and was widely studied because it shows some similarity with DNA unwinding in the cell.  There is recent renewed interest in biology due to high resolution melting methods \cite{RefHRM}.
  
Despite numerous attempts, the understanding of this transition is still a challenge.  One difficulty in making progress is the absence of structural information at the transition.  This is because experimental studies of DNA melting, made using techniques such as UV absorbance, circular dichroism and calorimetric studies, do not provide spatial information.  

Diffraction techniques can provide this information.  Indeed, the double-helix structure of DNA was solved by modeling the x-ray diffraction patterns from semi-crystalline fibre samples, measured by Franklin \emph{et al.} \cite{Franklin_a}.  The data were analyzed for peak positions which revealed the double helix structure, and also that the molecules could have different configurational structures \cite{Franklin_a}.  Many configurations are known to exist \cite{Fuller}, however the majority of the work to date has been on so-called `B-form' DNA. 

A Bragg peak contains more information than simply its position.  Analysis of the shape and width can determine a correlation function and its characteristic length.  The integrated intensity gives a measure of the quantity of the sample that scatters coherently.  These quantities change dramatically close to a phase transition, and scattering techniques have proved to be excellent probes  \cite{Cowley}.

We have used neutron scattering to measure the temperature dependence of a strong Bragg peak from a fibre sample of B-form DNA.  The data have been analyzed to extract the widths and the integrated intensities as the sample was heated through the melting transition.  The structural information was compared with calculations using an adapted form of a mesoscopic statistical mechanics model for DNA, known as the Peyrard-Bishop-Dauxois (PBD) model \cite{PBD}.

While the DNA molecules in a fibre are confined,  large scale configurational changes are possible and are driven by a simple change in relative humidity \cite{Franklin_a,Fuller}.  DNA melting is observable in a fibre and, given the favourable agreement between the theory and experiment, much of the physics of the transition can be understood from the application of the PBD model.

The fibre sample was created using the ``wet spinning" method \cite{Rupprecht} with DNA extracted from salmon testes (Fluka), precipitated from a 0.4 M Li-salt solution.  Samples produced in this method are semicrystalline \cite{Fuller} with the DNA molecules aligned to within $5^{\circ}$ of the fibre axis \cite{Grimm}.  The water content was set by keeping the sample in an atmosphere humidified to 75 \% with deuterated water.  This ensured a B-form configurational structure, proven with  X-ray fibre diffraction, and significantly reduced incoherent neutron scattering from protonated hydrogen in the sample.

\begin{figure}
\begin{center}
\includegraphics[width=8cm]{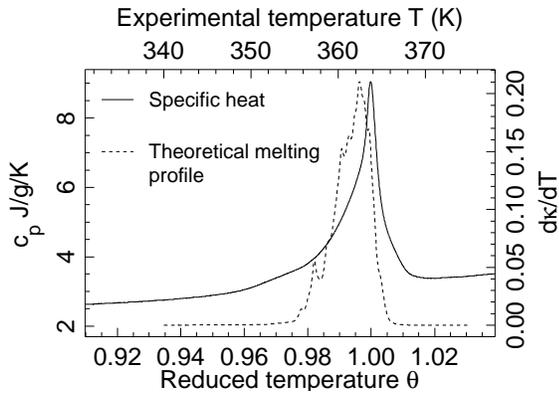}
\caption{Differential calorimetry signal, $C_p$, from a  Li-DNA  sample similar to that used in the neutron experiments, and the theoretical denaturation profile calculated from \emph{Pyrococcus abyssi} DNA used for the analysis where $\kappa$ is the fraction of open base pairs}
\label{fig.DSC} 
\end{center}
\end{figure}

The thermodynamic behaviour of B-form fibre DNA was characterized by measuring the heat capacity using differential scanning calorimetry (DSC) at the Institut N\'eel, Grenoble, France.  The samples were hermetically sealed in a hastalloy sample tube.  The specific heat was measured in a Seratam Micro DSC III calorimeter relative to an empty reference tube.  The differential heat flux was measured as the samples were heated from 278 K to the maximum temperature ($\sim 380$ K) at a rate of 0.6 K / min.   Typical data are shown in Fig. \ref{fig.DSC}.  The data are sharply peaked at 360 K for this sample, which represents its thermal denaturation temperature.  The transition is not reversible.

\begin{figure}
\begin{center}
\includegraphics[width=8cm]{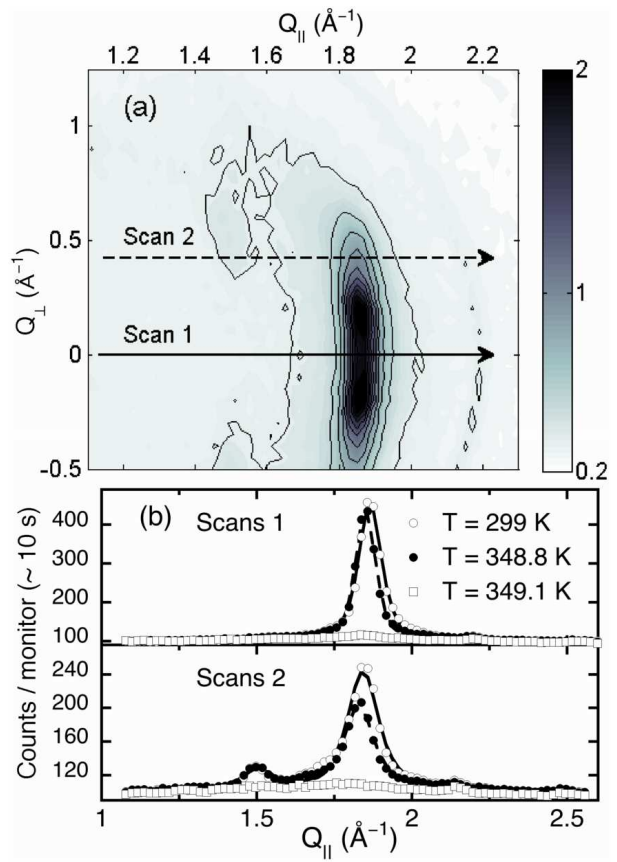}
\caption{a) Reciprocal space map of B-form Li-DNA.   The axes are the momentum transfer parallel ($Q_{\|}$) and perpendicular ($Q_{\perp}$) to the fibre axis.  The strong Bragg peak is observed at $Q_{\|} = 2\pi/a$, where $a \approx 3.4$ {\AA} is the distance between the base pairs along the fibre axis.  b) Examples of scans, fitted using eq. \ref{eq.SQw_exp}.  The data at 299 K represent the starting point for the experiment.  The sample is in the melting transition at 348.8 K.  The fibre structure has collapsed at 349.1 K.}
\label{fig.RMap}
\end{center}
\end{figure}

After humidification, the sample for the neutron scattering measurements was placed in a Nb envelope and then sealed in an Al cassette using Pb wire for the seal.  The cassette assisted in the thermalization of the sample and ensured that the water content for the sample environment stayed constant throughout the measurement.  The scattering was first characterized using the IN3 spectrometer at the Institut Laue-Langevin (ILL), France.  The main Bragg peak, found at a momentum transfer along the fibre axis, $Q_{\|}$, is seen in the reciprocal space map in Fig. \ref{fig.RMap}a.  Its lattice spacing corresponds to the average distance between the base pairs, $a \approx 3.4$ {\AA}.  The peak is very broad in the momentum transfer perpendicular to the fibre axis, $Q_{\perp}$.  The sample can largely be regarded as a quasi-1D crystal, and is treated as such.  Some DNA features are seen off the fibre axis due to its double helix structure, as is a weak powder ring from Pb at $|{\bf{Q}}| = 2.2$ {\AA}$^{-1}$.  Stronger rings from Al, seen at $|{\bf{Q}}| = 2.68$ {\AA}$^{-1}$, were monitored throughout the experiment as a check on instrument and sample alignment. 

The neutron three axis spectrometer IN8, also at ILL, was used to measure the main Bragg peak as a function of temperature.  This instrument was configured with an incident wavelength of 1.53 {\AA} ( $\equiv 35$ meV).  The Q-resolution was defined with 40$^{\prime}$ collimation before and after the sample, and was measured by making a reciprocal space map of a Bragg peak from a Si single crystal.  No energy analysis was used,  thus the static approximation was assumed to hold for the measurements.  Temperature control was achieved using a liquid helium cryofurnace.  

Two scans were repeated at all temperatures.  Examples are shown in Fig. \ref{fig.RMap}b. Their trajectories were calculated for nominally elastic scattering and are shown in Fig. \ref{fig.RMap}a.  Scan 1 was along the fibre axis, through the centre of the Bragg peak.  The $Q_{\perp}$ for scan 2 was chosen such that, when $Q_{\|} = 2\pi/a$, the direction of the scattered beam would be perpendicular to the fibre axis.  This type of scan has been used to measure critical phase transitions in low dimensional magnets \cite{Cowley}.  The temperature steps close to the melting transition were very small (0.1 K) and measurements at a given temperature were repeated numerous times to ensure thermal equilibrium and reproducibility. 

The data were fitted with a Lorentzian lineshape:
\begin{equation}
S\left( Q_{\|} \right) = \int^{\infty}_{-\infty}S\left(Q_{\|},\omega\right){\text d}\omega=\frac{I_0}{\pi}\frac{\Gamma / 2}{\left( \Gamma / 2 \right)^2 + \left (Q_{\|} - Q_0\right)^2},
\label{eq.SQw_exp}
\end{equation}
where $I_0$ is the integrated intensity, $Q_0$ is the peak centre, and $\Gamma$ is the width which is inversely proportional to the correlation length along the molecule.  The function was convoluted with the instrument resolution.  A second Lorentzian centered at $Q_{\|} \approx 1.5$ {\AA}$^{-1}$ was needed to fit the scan 2 data.  The amplitudes for these two peaks were free parameters, however their widths were set to be equal in the fits.  
Fig. \ref{fig.RMap}b shows examples of the fits.

\begin{figure}
\begin{center}
\includegraphics[width=8cm]{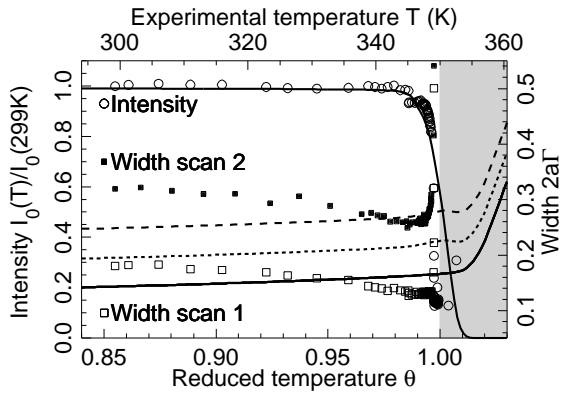}
\caption{The integrated intensities and widths of the Bragg peaks as a function of temperature.  The points show the experimental results, while the lines plot the theory.  For the widths: the solid line shows the calculation for scans 1; the dotted line for scans 2 assuming uncorrelated components of structural disorder ($\chi = 0$); the dashed line for scans 2 and correlated disorder with $\chi = 0.35$.}
\label{fig.AmpWid}
\end{center}
\end{figure}

The fit results are shown in Fig. \ref{fig.AmpWid}.  The normalized intensities for both scans followed exactly the same curve.  A small deviation in the data at T $\approx 345$ K was due to the discovery of an alignment issue that was found and rectified.  The widths were qualitatively different and are shown for both scans.

The samples used for the neutron and DSC measurements were produced by the same methods and apparatus with nominally the same ion concentration and humidity.  While melting transitions are qualitatively the same, the precise temperatures involved depend strongly on external conditions \cite{Frank}.  Hence, it is no surprise that the transition temperatures differ between neutrons and DSC.  Subsequent analysis will discuss the transition in terms of a reduced temperature, $\theta = T/T_c$, where $T_c$ is defined as where half of the base pairs are open. 

The theoretical modeling of the data required the derivation of the scattering profile from a finite length of DNA, accounting for disorder due to variations in the base pair sequence and to thermal fluctuations.  A generalization of the profile for a 1D paracrystal of finite size \cite{Mu} results in an expression for $m$ closed base pairs:
\begin{equation}
\begin{array}{l}
S_m\left({\bf Q}\right)=m+2\sum_{n=1}^{m-1}\left(m-n\right)\cos\left( Q_{\|}na\right)e^{-n\Delta}, \\
\lim_{m \rightarrow \infty}\left( S_m\left({\bf{Q}}\right)/m\right) \approx \mathcal{S}^{\prime}\left({\bf{Q}}\right)= \frac{\sinh\Delta}{\cosh\Delta - \cos\left(Q_{\|}a\right)},
\end{array}
\label{eq.SQ_1D}
\end{equation}
where $2\Delta = Q^2_{\|}\left(\left< \sigma^2_{\|}\right>+\left<\xi^2_{\|}\right>  \right)+Q^2_{\perp}\left( \left<\sigma^2_{\perp}\right>+\left<\xi^2_{\perp}\right>\right)+2Q_{\perp}Q_{\|}\chi\left(\left<\sigma^2_{\perp}\right>\left<\sigma^2_{\|}\right>\right)^{1/2}$.  The Gaussian displacements,  $\sigma_{\perp,\|}$, characterize the structural disorder perpendicular and parallel to the molecule, and may be obtained from conformational analysis \cite{Lavery} to be $\left<\sigma^2_{\|}\right>^{1/2}$= 0.18 {\AA} and $\left<\sigma^2_{\perp}\right>^{1/2}$ = 0.55 {\AA}.  The terms, $\xi_{\perp,\|}$ account for thermal fluctuations.  Along the molecule, $\xi_{\|}$ amounts to a Debye-Waller correction that can be calculated from the total DNA mass per base pair and from the measured sound speed \cite{Grimm, Krisch}, while $\xi_{\perp}$ approximates displacements perpendicular to the molecule axis.  The parameter $\chi$ characterizes the correlation between longitudinal and transverse structural disorder, with $\chi = 0$ for no correlation.  

The PBD model \cite{PBD} was developed to describe the melting behaviour of DNA.  It may be used to calculate the probability, $P\left(m\right)$, of having a closed cluster of $m$ consecutive base pairs at a given temperature.  It has been recently extended to model genomic DNA with no free parameters \cite{Theo1}, and appropriate parameters from that work were used in the calculations here.  The parameters were derived using sequence information from \emph{pyrococcus abyssi} DNA and were used to model the DSC data in Fig. \ref{fig.DSC} with satisfactory agreement.  Natural DNA has $\sim 10^6$ base pairs.  While $P\left(m\right)$ may be calculated up to $\sim 10^2$ base pairs, it will scale\cite{Theo2} exponentially with $m$ and may be replaced by $P\left(m > M\right) = P_0\zeta^{m}$.  The expected neutron scattering can then be calculated in three regimes using the equation:
\begin{equation}
\begin{array}{c}
S_C({\bf{Q}}) =  \sum^{M}_{m=1} P(m) S_m({\bf{Q}}) 
+  \sum_{m=M+1}^{M_0} P_0 \zeta^m S_m({\bf{Q}}) \\
+ \mathcal{S}^{\prime}({\bf{Q}}) P_0 \zeta^{M_0 + 1} \left[ \frac{M_0}{1 - \zeta} +  \frac{1}{(1 - \zeta)^2} \right].
\end{array}
\label{eq.SQ_theor}
\end{equation}
The first term was calculated for $M = 150$ and the others for $M_0 = 1000$.  The calculated structure factors were then fitted with eq. \ref{eq.SQw_exp}.  Fig. \ref{fig.AmpWid} shows the results, and the experimental data, as a function of $\theta$.

Theory and experiment are very well matched until $\theta \approx 1$, in particular for the integrated intensities where the agreement is almost perfect.  As shown by eq. \ref{eq.SQ_theor}, the intensity scales with the number of closed pairs.  The intensities are almost constant with increasing temperature until $\theta \approx 0.97$, due to there being a very low fraction of open base pairs in this temperature range.   They then begin to fall in a smooth manner as the number of closed pairs decreases.  This rounded, continuous behavior of the transition over a finite temperature range represents the ``genomic" limit of multistep melting proposed in model calculations of heterogeneous DNA \cite{Hwa}.

The comparison of the widths provide spatial information that standard observations of DNA denaturation cannot.  Globally, the comparison between theory and experiment is reasonable.  For $\theta < 0.97$ there is a qualitative difference, with the calculated widths slowly increasing with temperature while the experimental data show a decrease.  Subsequent experiments show that this is a form of annealing in the sample, with the widths staying constant when the temperature is decreased, and hence is due to the fibre structure rather than DNA denaturing.

There are no free parameters in the calculation of the widths for scan 1, where $Q_{\perp} = 0$.  Theory and data match quantitatively with a discrepancy of $< 15 \%$, validating the conformational analysis \cite{Lavery}. The widths hardly vary with temperature until $\theta = 1$, showing that the base pair openings, which start to be significant at $\theta = 0.98$, do not cause a sharp decrease in the size of the diffracting clusters until denaturation has occurred.  Clusters of $\sim 100$ base pairs remain intact well in to the denaturation region and then denature as a whole.

Scans 2, with $Q_{\perp} \ne 0$, require the introduction of $\chi$.  Two calculations are shown in Fig. \ref{fig.AmpWid}.  The first assumes no correlation ($\chi = 0$) and is unacceptably lower than the experimental data.  A moderate value ($\chi = 0.35$) gives a width that matches the data for $\theta \approx 0.97$, judged to be representative for an ``annealed" sample.  It would be interesting to test the value of $\chi$ with conformational analysis. 

The experimental widths of scans 2 show an upturn in the vicinity of the phase transition ($0.99 \le \theta < 1$) which is not accurately modeled.  The theory does account for transverse thermal fluctuations through $ \left<\xi^2_{\perp}\right> $, however this effect is small as only fluctuations from closed clusters can contribute.  More likely, the upturn is due to thermally induced misalignment of the molecules with respect to the fibre axis.  A misalignment of $10 \%$ would change the width along the fibre axis by $\delta Q_{\|}/Q_{\|} = 2 \%$, but the projection at non-zero $Q_{\perp}$ would create a change of $17 \%$.  The assumption of a 1D structure begins to break down at these temperatures and the theory would have to be extended to account for that.

The theory and experiment cease to match at $\theta \approx 1$ when the experimental intensities fall precipitously while the theory continue to show a narrow but smooth decay.  The data in Fig. \ref{fig.RMap}b show examples of the scattering below and above this temperature.  The fall is due to the collapse of the fibre structure, as verified by optical microscopy on similar samples and by visual inspection of this sample after the experiment.  The structural change in the fibre is distinct from DNA melting, which refers to the denaturing of the double helix, and is not explicitly described by the theory.  Nevertheless, the theory gives some insight to the phenomenon.  As $T_c$ is approached the open regions will quickly grow in size.  Single strands are very flexible and can gain entropy by losing their initial orientation.  The resulting structure would have the remaining closed, relatively rigid regions embedded in a liquid-like medium.  Once the open regions reach a certain average size, the closed regions will have flexibility to move.  The highly oriented fibre structure will break down, heralded by the upturn in the scan 2 widths and the disappearance of the Bragg peak seen in Fig. \ref{fig.RMap}b.

Despite the loss of the Bragg peak, it is interesting to note that the theory predicts that the size of the closed segments is still large ($\sim 80 - 90$ base pairs) slightly above $\theta = 1$.  This is indirectly confirmed by agarose gel electrophoresis analysis before and after the experiment.  Before heating, the DNA had $> 20000$ base pairs, while after it had broken in to fragments of $\sim 100$ base pairs.  The breaking can be due to the high stress concentration that occurs at the end of the rigid closed segments, linked to each other by flexible single strands, as they rotate in the breakdown of the fibre structure.

In conclusion, neutron scattering has been used to give spatial information on the melting of B-form DNA in fibre form.  The data could be understood using simple non-linear lattice dynamics theory with minimal free parameters, showing the transition to be smooth and continuous.  The comparison was excellent to a point mid-way through the melting transition where the analysis was complicated by the collapse of the fibre structure.  Results showed that large clusters of closed base pairs continue to exist well in to the denaturation regime, and showed correlations between thermal fluctuations along and perpendicular to the molecule.

The authors would like to thank Prof. M. Johnson, Dr. L. van Eijck, and Prof. V. T. Forsyth for valuable discussions.  The allocation of beam time by the ILL and the aid of the IN8 and IN3 instrument teams are gratefully acknowledged.

\bibliographystyle{unsrt}  

\begin{thebibliography}{10}

\bibitem{Zimm}
B. H. Zimm and J. K. Bragg, J. Chem. Phys. {\bf 31}, 526 (1959)

\bibitem{Dauxois}
T. Dauxois, N. Theodorakopoulos and M. Peyrard, J. Stat. Phys. {\bf 107}, 869 (2002) 

\bibitem{Thomas}
R. Thomas, Biochimica et Biophysica Acta {\bf 14}, 231 
(1954)

\bibitem{Rice}
S.A. Rice and P. Doty, J. Am. Chem. Soc. {\bf 79}, 3937 
(1957)

\bibitem{RefHRM}
C.T. Wittwer, Human Mutation {\bf 30}, 857-859 (2009)

\bibitem{Franklin_a} R. E. Franklin and R. G. Gosling, Acta Cryst. {\bf 6}, 673 (1953) 

\bibitem{Fuller} W. Fuller, T. Forsyth and A. Mahendrasingam, Phil Trans. R. Soc. Lond B {\bf 359} 1237 (2004).

\bibitem{Cowley} R. A. Cowley, in \emph{Methods of Experimental Physics} vol. 23 part C. K. Sk\"old and D. L. Price (eds)  (Academic, Orlando, 1987) pp. 1 - 68

\bibitem{PBD}
 M. Peyrard, A.R. Bishop, Phys. Rev. Lett. {\bf 62}, 2755 (1989); T. Dauxois, M. Peyrard and A.R. Bishop, Phys. Rev. E {\bf 47}, R44 (1993).

\bibitem{Rupprecht}
A. Rupprecht, Acta Chem. Scan. {\bf 20} 494 (1966); A. Rupprecht, Biotech. and Bioeng.  {\bf XII} 93 (1970)

\bibitem{Grimm}
H. Grimm \emph{et al.}, Phys. Rev. Lett. {\bf 59} 1780 (1987).

\bibitem{Frank}
M. D. Frank-Kamenetiskii, Biopolymers {\bf 10} 2623 (1971); M. D. Frank-Kamenetiskii, Phys. Rep. {\bf 288} 14 (1997).

\bibitem{Lavery}
R. Lavery \emph{et al.}, Nuc. Acis. Res. {\bf 37} 5917 (2009).

\bibitem{Krisch}
M. Krisch \emph{et al.}, Phys. Rev. E {\bf 73} 061909 (2006).
 
\bibitem{Mu}
X.-Q. Mu, Acta Cryst. A {\bf 54} 606 (1998); M. Peyrard {\it et al.}, in preparation

\bibitem{Theo1}
N. Theodorakopoulos, Phys. Rev. E {\bf 82} 021905 (2010).

\bibitem{Theo2}
N. Theodorakopoulos, Phys. Rev. E {\bf 77} 031919 (2008).

\bibitem{Hwa} D. Cule and T. Hwa,
Phys. Rev. Lett. {\bf 79}, 2375
(1997).





\end{thebibliography}

\end{document}